\documentclass[aip,jcp,preprint,amsmath]{revtex4-1}
\usepackage[bookmarks,bookmarksopen]{hyperref}
\usepackage{graphicx} 
\usepackage{epsfig}
\usepackage{dcolumn}
\usepackage{bm}
\usepackage{longtable}

\newcommand{\az}{$a_{0}$}
\newcommand{\cm}{cm$^{-1}$}
\newcommand{\dg}{$^\circ$}

\newcommand{\mnras}{Mon. Not. R. Astron. Soc.}
\newcommand{\apj}{Astrophys. J.}
\newcommand{\apjl}{Astrophys. J. Lett.}
\newcommand{\jcp}{J. Chem. Phys.}

\begin{document}

\title{The interaction of He with vibrating HCN: potential energy surface, bound states and rotationally inelastic cross sections}

\author{Otoniel Denis-Alpizar}
\email{otonieldenisalpizar@gmail.com} 
\affiliation{Universit\'e de Bordeaux, ISM, CNRS UMR 5255, 33405 Talence Cedex, France}
\affiliation{Departamento de F\'isica, Universidad de Matanzas, Matanzas 40100, Cuba} 
\author{Thierry Stoecklin} 
\author{Philippe Halvick} 
\affiliation{Universit\'e de Bordeaux, ISM, CNRS UMR 5255, 33405 Talence Cedex, France}
\author{Marie-Lise Dubernet}
\affiliation{Universit\'e Pierre et Marie Curie, LPMAA, UMR CNRS 7092, 75252 Paris, France}
\affiliation{Observatoire de Paris, LUTH, UMR CNRS 8102, 92195 Meudon, France}

\date{\today}

\date{\today}

\begin{abstract}
A four-dimensional potential energy surface representing the interaction between
He and HCN subjected to bending vibrational motion is presented. \textit{Ab initio}
calculations were carried out at the coupled-cluster level with single
and double excitations and a perturbative treatment of triple excitations,
using a quadruple-zeta basis set and mid-bond functions. The global minimum is found in the linear He-HCN configuration with the H atom pointing towards helium at the intermolecular
separation of 7.94 a$_{0}$. The corresponding well depth is 30.35~cm$^{-1}$.
First, the quality of the new potential has been tested by performing two
comparisons with previous theoretical and experimental works. \textit{i})
the rovibrational energy levels of the He-HCN complex for a rigid linear
configuration of the HCN molecule have been calculated. The dissociation
energy is 8.99~\cm, which is slightly smaller than the semi-empirical
value of 9.42~\cm. The transitions frequencies are found to be in
good agreement with the experimental data. \textit{ii}) we performed close
coupling calculations of the rotational de-excitation of rigid linear HCN in collisions
with He and observed a close similarity with the theoretical data
published in a recent study. Second, the effects of the vibrational bending of HCN have been
investigated, both for the bound levels of the He-HCN system and for the rotationally inelastic cross sections. This was performed with an approximate method using the average of the interaction potential over the vibrational bending wavefunction. While this improves slightly the comparison of calculated transitions frequencies with experiment, the cross sections remains very close to those obtained with rigid linear HCN. 
\end{abstract}

\maketitle

\section{Introduction}

The rigid monomer approximation (RMA), where the monomer's geometry
is assumed to be independent of the dimer configuration, is commonly
used to simulate the dynamics of systems governed by weak intermolecular
interaction and where no breaking or formation of chemical bonds take
place. The decoupling of intramonomer and intermonomer motions reduces
the dimensionality and thus simplify greatly the calculation of the
dynamics. Intermolecular bound states or cross sections for low collision
energies can be calculated within this approximation. The quality
of the approximation can be improved by using the average of the intramonomer
coordinates over the internal stretching motions\cite{jeziorska:00}.
When the ratio of intramonomer over intermonomer vibrational frequencies
is large (about 100), the RMA is very reliable. This has been demonstrated\cite{jankowski:05}
by the excellent agreement between experiment and calculation of the
infrared spectrum of the H$_{2}$-CO complex. For the same system,
a good agreement has been also obtained between calculations and the
first low temperature experimental inelastic cross section\cite{chefdeville:12}.
However, in the case of a triatomic (or larger) monomer, the RMA can
be questioned because the coupling between the internal bending motion
and the intermonomer motion may not be negligible. Bending motion
may have large amplitude and low frequency, inducing a significant
change of the electronic cloud, and consequently, a significant change
of the intermolecular forces. While the RMA should be useless for very floppy monomer (\textit{e.g.} C$_3$), it is not known if this method can be accurate for rigid or semi-rigid molecules with vibrational bending mode.


Hydrogen cyanide (HCN) and isocyanide (HNC) are among the most abundant
organic molecules in the interstellar medium. Owing to a large dipole
moment, both molecules decay fast in their rotational energy ladder.
The rotational emission lines of HCN and HNC are considered to be
a major tracer of dense molecular gas (star-forming molecular clouds)
in luminous and ultraluminous infrared galaxies \cite{gao:04,gao:07,gracia-carpio:08,baan:08}.
Rotationally excited HCN and HNC suggests an excitation mechanism
fast enough to counter the decay, such as frequent collisions with
He and H$_{2}$ in dense clouds. Consequently, the estimation of abundances
of both isomers in the interstellar clouds has motivated theoretical
studies of the rotational excitation in collisions with He\cite{green:74,monteiro:86,sarrasin:10,dumouchel:10}
and H$_{2}$\cite{dumouchel:11}. In these studies, the HCN or HNC molecule
was always considered as a linear rigid rotor.

However, vibrational excitation of HCN has been observed in the interstellar
medium. The rotational transitions of vibrationally excited HCN have
been used to probe\cite{cernicharo:11} the dust formation region
around the carbon-rich star IRC +10216. The high vibrational levels
are populated by radiation and by collision, owing to the high temperature,
high gas density and high radiation flux prevailing in the circumstellar
envelope. Vibrationally excited HCN in the $\nu_{2}$=1 state has
been also observed\cite{sakamoto:10} in the nucleus of the luminous
infrared galaxy NGC 4418. Most likely, the molecule is pumped to the
excited level by infrared radiation and return to the vibrational
ground state with rotational excitation\cite{morris:75,carroll:81}.
These observations suggest that the vibrational excitation of HCN,
at least in the bending motion, deserves to be considered in the collision
mechanisms.

The first studies dedicated to the rotational excitation of rigid linear HCN
($l$-HCN) by collisions with He atoms were based on the potential
energy surface (PES) of Green and Thaddeus\cite{green:74}. This primitive
PES was obtained using the uniform electron gas model. Several new
intermolecular potentials were later published in the last twenty years
for the $l$-HCN -- He system. Drucker \textit{et al.}\cite{drucker:95}
calculated one at the MP4 level and reported the first theoretical
determination of the high-resolution microwave and millimeter spectrum.
Later Atkins and Hutson\cite{atkins:96} obtained two empirical PESs
based on two different functional forms using the experimental data
available. Toczylowski \textit{et al.}\cite{toczylowski:01} reported
a theoretical PES calculated at the CCSD(T) level, hereafter denoted
by S01, which was found to describe correctly the internal-rotational
band measured by Drucker \textit{et al.} and with a global minimum
of -29.90~cm$^{-1}$. The most recent studies of the rotational excitation
of $l$-HCN by He done by Sarrasin \textit{et al.}\cite{sarrasin:10}
and Dumouchel \textit{et al.}\cite{dumouchel:10} used this last surface.
The latest PES published for the $l$-HCN -- He system is a semi-empirical
one by Harada \textit{et al.}\cite{harada:02} denoted S02, which
was obtained by modifying the S01 surface in order to reproduce the
experimental transitions frequencies.

The present paper focus on the development of a PES describing the
collision between He and HCN considered as a rigid bender. The vibrational
bending motion of HCN is treated quantally while the CH and CN bond
lengths are set to constant values. As a first test of this new PES,
we determined the rovibrational energy levels of the $l$-HCN -- He
system and compared it to the existing theoretical and experimental
data. We also computed the $l$-HCN -- He inelastic cross sections
and compared it with the theoretical data of Sarrasin \textit{et al.}
In the second part of this work, the effects of the vibrational bending of HCN have been investigated by using an interaction potential averaged on the bending wavefunctions\cite{valiron:08,lifangma:12}.
Again, we calculated the energies of the rovibrational bound states and the inelastic cross sections and we compared these last results with the previous ones.

\section{ Ab initio calculations and potential functional form}

\begin{figure}[!!ht]
 \centering \includegraphics[width=0.3\textwidth]{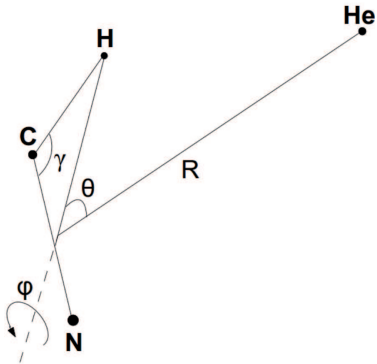} 
\caption{Definition of the body-fixed coordinate system for the He - HCN system.
The planar configuration represented here corresponds to $\varphi$=180\dg.
The angle $\varphi$ is not defined for $\gamma$ or $\theta$ equal
to 0\dg or 180\dg. }
\label{coord} 
\end{figure}

The body-fixed coordinates used in this work are shown in Fig. \ref{coord}.
$R$, $\theta$ and $\varphi$ are the intermonomer coordinates which
describe the relative positions of the HCN molecule and He atom, while
$\gamma$ is the intramonomer coordinate which describes the bending
angle of HCN. $R$ is the distance from the center of mass of the
HCN to the He atom. $\varphi$ is the angle of rotation around the
axis defined by the H atom and the center of mass of CN. $\theta$
is the angle between the latter axis and the axis defined by the He
atom and the center of mass of HCN. The C-H and C-N rigid bond lengths
have been fixed to the sum of the experimental value \cite{strey:73}
plus the correction for the averaging over the ground vibrational
state \cite{laurie:62}, which results to r$_{CH}$ = 2.0286~\az
\ and r$_{CN}$ = 2.1874~\az.

%

The interaction potential of HCN with He has been calculated in the
framework of the supermolecular approach with the coupled-cluster
method with single and double excitations and a perturbative treatment
of triple excitations (CCSD(T)). The interaction energy was corrected
at all geometries for the basis set superposition error (BSSE) with
the counterpoise procedure of Boys and Bernardi\cite{boys:70}. A
comparison of the interaction energies calculated with basis sets\cite{woon:93}
of triple, quadruple and quintuple-zeta quality is shown in table
\ref{tab:eint}, with or without an additional set of bond functions
\cite{cybulski:99} centered at mid-distance between the He atom and
the HCN center of mass. The interaction energy, calculated at a configuration
close to the equilibrium geometry, is quite stable in respect of the
size of the basis set and the use of bond functions. For the largest
basis set, it is safe to assume that the convergence of the one-electron
basis is close to the complete basis set limit. Considering the computational
cost associated with the various basis sets, we have chosen the quadruple
zeta basis set with bond functions.


\begin{table}
\caption{\label{tab:eint} CCSD(T) interaction energy of the $l$-HCN-He
system at $R$ = 7.97~\az\ and $\theta$ = 0\dg. The use of bond functions is denoted
by +bf}
\centering{}\setlength{\tabcolsep}{13pt} \begin{tabular}{lcc}
\hline 
Basis set  & Energy (\cm)  & Relative computational cost \tabularnewline
\hline 
aug-cc-pVTZ+bf  & -29.85  & 1 \tabularnewline
aug-cc-pVQZ  & -29.64  & 3.3 \tabularnewline
aug-cc-pVQZ+bf  & -30.34  & 6.2 \tabularnewline
aug-cc-pV5Z  & -30.28  & 21.5\tabularnewline
\hline
\end{tabular}
\end{table}

The interaction energy was computed over a dense four-dimensional
grid of points defined by the product of four one-dimensional grids
associated to a single coordinate. The radial grid included 35 points
ranging from 3.8~\az \ to 20.8~\az. The bending grid included
11 points between 180\dg \ and 110\dg. The angular grids were spaced
uniformly in steps of 10\dg \ for $\theta$ and 30\dg \ for $\varphi$,
both in the range {[}0\dg,180\dg{]}. The total number of points
was 43015. All calculations were carried
out with the Molpro package \cite{molpro2010}.

The \textit{ab initio} energies were fitted to a parametrized functional form
defined as a sum of a short-range and a long-range contributions:
\begin{equation}
\begin{split}V_{int}(R,\theta,\varphi,\gamma)= & \, S(R)\sum_{l=0}^{14}\sum_{m=0}^{\min{(l,3)}}F_{lm}^{{\scriptscriptstyle SR}}(R,\gamma)\bar{P}_{lm}(\theta)\cos{(m\varphi)}\\
 & +(1-S(R))\sum_{l=0}^{5}\sum_{m=0}^{\min{(l,3)}}F_{lm}^{{\scriptscriptstyle LR}}(R,\gamma)\bar{P}_{lm}(\theta)\cos{(m\varphi)}\end{split}
\end{equation}

Here, $\bar{P}_{lm}$ are normalized associated Legendre polynomials.
$F_{lm}^{{\scriptscriptstyle SR}}$, $F_{lm}^{{\scriptscriptstyle LR}}$
and $S$ are the short-range radial functions, the long-range radial
functions and the switching function respectively: \begin{equation}
F_{lm}^{{\scriptscriptstyle SR}}(R,\gamma)=e^{-\alpha R}\sum_{n=0}^{9}R^{n}\sum_{j=0}^{3}C_{lmnj}\bar{P}_{j}(\cos{\gamma})\end{equation}
 \begin{equation}
F_{lm}^{{\scriptscriptstyle LR}}(R,\gamma)=\sum_{k=6}^{8}\frac{t_{k}(\beta R)}{R^{k}}\sum_{j=0}^{3}D_{lmkj}\bar{P}_{j}(\cos{\gamma})\end{equation}
 \begin{equation}
S(R)=\frac{1}{2}[1-\tanh{(A_{0}(R-R_{0}))}]\end{equation}

where $\bar{P_{j}}$ are normalized Legendre Polynomials and $t_{k}$
is the Tang-Toennies damping function: \begin{equation}
t_{k}(x)=1-e^{-x}\sum_{i=0}^{k}\frac{x^{i}}{i!}\end{equation}

The non-linear parameters $\alpha$, $\beta$, $A_{0}$, and $R_{0}$
were set to the values $\alpha$ = 1.91~\az$^{-1}$, $\beta$ =
1.06~\az$^{-1}$, $A_{0}$ = 1.69~\az$^{-1}$ and $R_{0}$ = 10.58~\az.
The linear parameters $C_{lmnj}$ and $D_{lmkj}$ were calculated
with the weighted linear least squares method. On each \textit{ab
initio} point, we applied a weight $w$ depending both of the interaction
energy $E$ and the angle $\gamma$: \begin{equation}
w=\frac{\gamma_{0}}{(\tau-\gamma)^{2}}\min(1,\frac{V_{0}}{\mid E\mid})\end{equation}
 with $V_{0}$ = 1000~cm$^{-1}$, $\gamma_{0}=100^{\circ}$ and $\tau=181^{\circ}$.

Let us note that the \textit{ab initio} grid is restricted to $\gamma\geq110^{\circ}$.
Indeed, the rigid bender approximation used for HCN is expected to
be reliable only for the ground and the first excited bending states,
and possibly for the second excited state. Moreover, the potential
energy of the HCN molecule at $\gamma=120^{\circ}$ is 7130~cm$^{-1}$.
This value is much larger than the energy at which the rigid bender
approximation remain reliable, if we remind that $\omega_{2}$ is
slightly larger than 700~cm$^{-1}$. Therefore, because there is
no need to represent the interaction energy for $\gamma\leq120^{\circ}$,
this value is used as a cut-off limit. Below this limit, the interaction
energy is set equal to its value at $\gamma=120^{\circ}$.

The total PES is the sum of the interaction energy
of the He - HCN complex plus the bending energy of the isolated HCN
molecule. The latter was calculated with the same \textit{ab initio}
method and same basis set which were used for the former. A grid of 22 points
was calculated and fitted to a linear combination of four Legendre
polynomials.

\section{Bound states and scattering calculations}

We used the close coupling method to calculate both the rovibrational
energy levels and the inelastic cross section of the He - HCN system.
The coupled equations needed for scattering calculations are identical
to those for bound states, the only difference being the applied boundary
conditions. In this study we compare two approaches. In the
first one, the bending motion is completely neglected and we use only
the linear configuration of HCN and perform usual atom linear molecule
calculations using for HCN a rigid rotor description. In the second
one, we fix the value of $\varphi$ to 0 as the potential varies
slowly as a function of this angle and we calculate for each value
of the intermolecular coordinate $R$ used in the dynamics calculations
the following expansion of the interaction potential in a Legendre polynomial
$P_{l}$(cos$\theta$) basis set along a grid of the bending angle $\gamma$:

\begin{equation}
V_{int}(R,\theta,\varphi=0,\gamma)=\sum_{l}D_{l}\left(R,\gamma\right)P_{l}(\cos\theta)
\label{eq:pot_exp}
\end{equation}

We then calculate the rigid bender energies and wavefunctions of
HCN in internal coordinates using the bending potential of HCN described
in the previous section and the Hamiltonian of Carter and Handy\cite{Carter:82}:

\begin{equation}
\begin{split}     
H_{RB}^{J=0}= & \,-\frac{\hbar^{2}}{2}\left[\frac{1}{\mu_{1}R_{1}^{2}}+\frac{1}{\mu_{2}R_{2}^{2}}\right]\left[\frac{\partial^{2}}{\partial\theta^{2}}+\cot\theta\frac{\partial}{\partial\theta}\right]\\
& -\frac{\hbar^{2}}{2M_{C}R_{1}R_{2}}\left\{ \left[\frac{\partial^{2}}{\partial\theta^{2}}+\cot\theta\frac{\partial}{\partial\theta}\right]\cos\theta
 +\cos\theta\left[\frac{\partial^{2}}{\partial\theta^{2}}+\cot\theta\frac{\partial}{\partial\theta}\right]\right\} \\
 &+\sum_{l}C_{l}P_{l}\left(\cos\theta\right)
\end{split}
\end{equation}

where $\frac{1}{\mu_{1}}=\frac{1}{M_{H}}+\frac{1}{M_{C}}$, $\frac{1}{\mu_{2}}=\frac{1}{M_{C}}+\frac{1}{M_{N}}$, and $R_{1}$ and $R_{2}$ are respectively the CH and CN bond lengths

It may be confusing to compare this Hamiltonian with the different
Hamiltonians published at that time\cite{Carter:82,Carter:83,Sutcliffe:82} 
as some other terms are present in some
of the three references (sometime with different signs) and are not in others. This
is probably because the matrix elements of these missing terms in
a Legendre polynomial basis set do compensate each other. The matrix
elements of this rigid bender Hamiltonian in a Legendre polynomial basis set
are:

\begin{equation}
\begin{split} 
\left\langle P_{l}\left|H_{RB}^{J=0}\right|P_{k}\right\rangle =& \,\frac{\hbar^{2}}{2}\left[\frac{1}{\mu_{1}R_{1}^{2}}+\frac{1}{\mu_{2}R_{2}^{2}}\right]\delta_{kl}\frac{2l\left(l+1\right)}{\left(2l+1\right)}-\frac{\hbar^{2}}{2M_{C}R_{1}R_{2}}\left[\frac{2l_{>}^{3}\delta_{k,l\pm1}}{\left(2l_{>}+1\right)\left(2l_{>}-1\right)}\right]\\
&+2\sum_{n} C_{n}\left(\begin{array}{ccc}
l & n & k\\
0 & 0 & 0\end{array}\right)^{2}
\end{split}
\end{equation}

where $l_{>}$~=~max($l$,$k$). The diagonalisation of this
matrix gives the rigid bender energies $\epsilon_{n}$ and wave functions
$\chi_{n}(\gamma)$ as a function of the bending angle for
the HCN rotational angular momentum $j=0$. We take the same
bending wavefunctions for all the values of $j$ since the variation
of the bending wavefunctions as a function of $j$ is expected to be weak, 
at least when $j$ is not too large. The
wavefunctions describing the HCN motion within this very simple approach are
then the product of a bending wavefunction by a spherical harmonics
describing the rotation. Consequently, the energies of HCN are:

\begin{equation}
E_{nj}=B_{HCN}j\left(j+1\right)+\epsilon_{n}
\end{equation}

The coefficients calculated in (\ref{eq:pot_exp}) are then averaged over the bending
wavefunctions:

\begin{equation}
\begin{split} 
\left\langle \chi_{n}\left|V_{int}(\bar{R},\theta,\varphi=0,\gamma)\right|\chi_{m}\right\rangle &= \sum_{l}\left[\int d\gamma\left\{ \chi_{n}\left(\gamma\right)D_{l}\left(\bar{R},\gamma\right)\chi_{n}\left(\gamma\right)\right\} \right]P_{l}(\cos\theta)\\
&=\sum_{l}\tilde{D}_{l}^{n,m}\left(\bar{R}\right)P_{l}(\cos\theta)
\end{split}
\end{equation}

The problem is now formally equivalent to an atom colliding a
fictitious vibrating diatomic molecule where the vibration of
the diatomic molecule is in fact the bending vibration. Using this
very simple approach denoted in the following Rigid Bender Averaged
Approximation (RBAA), we can obtain state to state cross sections
for the transition between two different bending and rotational levels
of HCN as well as bending averaged energies for the He-HCN complex.
We use our \begin{scshape}NEWMAT\end{scshape} code both for the scattering and the bound states
calculations. This is a close coupling code working in the spaced
fixed frame which has been described in some of our recent works\cite{halvick:11,lique:12}.

The rotational basis set for
HCN included 20 functions and the rotational constant of HCN was set
to its experimental value\cite{herzberg} B$_{HCN}$~=~1.47822~\cm.
The maximum propagation distance was 80~$a_{0}$ and two values of
the propagator step size (0.05~\az\ and 0.01~\az) were used for the bound
state calculations. The final bound state energies of the He-HCN complex
were obtained from a Richardson extrapolation. 


\section{Results and discussion}

\subsection{Potential energy surface}

The functional form defined above allowed us to obtain an accurate
representation of the PES. The root mean square (RMS) of the differences
between the \textit{ab initio} and the interpolated total potential energies $E$ is 0.016~\cm\ for the energies $E\leq$~0~\cm\ ($E$~=~0~\cm\ corresponds to the infinite separation of monomers). For 0~$<E\leq$~1000~\cm , the RMS of the relative errors is below 1\%, and for 1000~$<E\leq$~3000~\cm , is it about 2\%.

\begin{figure}[ht!!]
\centering{\includegraphics[width=10cm]{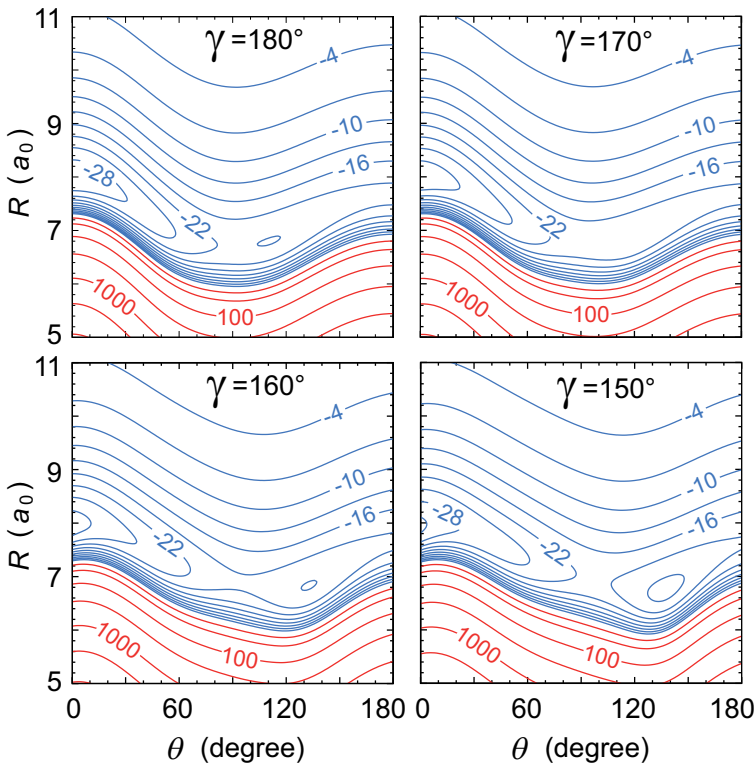}} 
\caption{Contour plot of the PES for selected values of $\gamma$ and for $\varphi$
= 0\dg. Negative contour lines (blue) are equally spaced by 3~\cm.}
\label{fig:phi000} 
\end{figure}

\begin{figure}[ht!!]
\centering{\includegraphics[width=10cm]{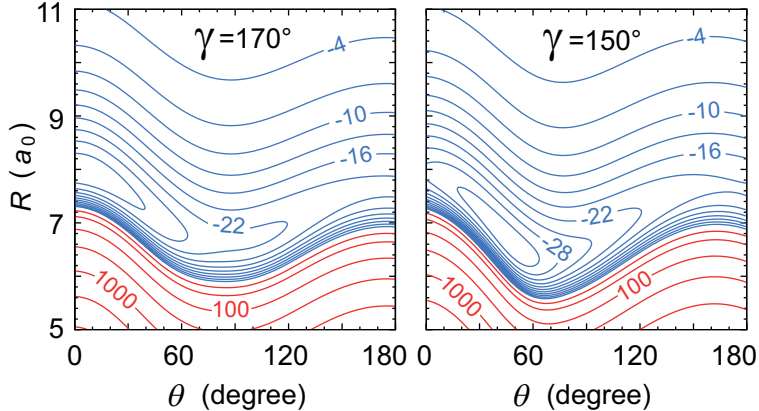}} 
\caption{Contour plot of the PES for selected values of $\gamma$ and for $\varphi$
= 180\dg. Negative contour lines (blue) are equally spaced by 3~\cm.}
\label{fig:phi180} 
\end{figure}

Contour plots of the interaction PES are shown in Fig. \ref{fig:phi000}
for several values of the bending angle $\gamma$ and for $\varphi$
fixed at 0\dg. The selected values of $\gamma$ are lying in the
range assumed to be spanned by the first excited vibrational function.
We observe that the bending of HCN has a visible effect in the
bottom of the potential well and in the repulsive short-range interaction. The long-range part  of the potential is hardly changed by the bending. For $\gamma$
= 180\dg, the potential is, by definition, isotropic versus $\varphi$.
In the range $150^{\circ}\leq\gamma\leq180^{\circ}$, the potential
remain nearly isotropic in respect of $\varphi$ (not shown here),
except in the short-range repulsive region. Contour plots for $\varphi$
= 180$^{\circ}$ are shown in Fig. \ref{fig:phi180}. For the same
value of $\gamma$, the comparison with the contour plots at $\varphi$
= 0$^{\circ}$ shows a significant change only for $\gamma$ = 150$^{\circ}$
and $R\leq$ 7~\az.

The global minimum of the total PES has a depth of 30.35~\cm\ and
a linear He-HCN configuration: $\gamma$ = 180$^{\circ}$, $\theta$
= 0$^{\circ}$ and $R$ = 7.94~\az. 
It is 0.45~\cm\ deeper than for the S01 PES. The latter
was calculated with a triple-zeta basis set, while we use here a quadruple-zeta
one. The discrepancy observed in the well depth is mainly a consequence
of the basis set quality, and this is confirmed by the data shown
in table I. The difference due to the different analytical representations
is probably not significant. Moreover, the present PES has a well
depth only 0.15~\cm\ larger than the one of the semi-empirical
surface S02, which is a S01 PES modified in order to improve the agreement
with the experimental millimeter-wave spectrum. A secondary minimum with a depth
of 22.08~\cm\ and a bent configuration is found at $\gamma$
= 180$^{\circ}$, $\theta$ = 110.4$^{\circ}$ and $R$ = 6.78~\az. This secondary minimum
is very similar to the global minimum of the He-CN PES\cite{lique:10CN}.

\subsection{Bound states and spectrum}

\begin{table}[ht!!]
\caption{\label{tab:bound} Bound levels of the He - HCN van der Waals complex.}
\centering{}\setlength{\tabcolsep}{13pt} \scalebox{0.93}{ \begin{tabular}{cccccc}
\hline 
 & \multicolumn{3}{c}{State} & RMA  & RBAA\tabularnewline
\hline 
 & $l$  & $J$  & $\varepsilon$  & Energy~(cm$^{-1}$) & Energy~(cm$^{-1}$) \tabularnewline
\hline 
$\nu_{s}$=0, $j$=0  &  &  &  &  & \tabularnewline
 & 0  & 0  & +  & -8.986  & -8.859 \tabularnewline
 & 1  & 1  & -  & -8.463  & -8.337 \tabularnewline
 & 2  & 2  & +  & -7.434  & -7.307\tabularnewline
 & 3  & 3  & -  & -5.928  & -5.801\tabularnewline
 & 4  & 4  & +  & -3.992  & -3.865\tabularnewline
 & 5  & 5  & -  & -1.676  & -1.550\tabularnewline
$\nu_{s}$=0, $j$=1  &  &  &  &  & \tabularnewline
 & 0  & 1  & -  & -5.619  & -5.515 \tabularnewline
 & 1  & 0  & +  & -5.207  & -5.097 \tabularnewline
 &  & 1  & +  & -5.089  & -4.986 \tabularnewline
 &  & 2  & +  & -5.004  & -4.905 \tabularnewline
 & 2  & 1  & -  & -4.153  & -4.046 \tabularnewline
 &  & 2  & -  & -3.954  & -3.855 \tabularnewline
 &  & 3  & -  & -3.822  & -3.730 \tabularnewline
 & 3  & 2  & +  & -2.588  & -2.485 \tabularnewline
 &  & 3  & +  & -2.278  & -2.186 \tabularnewline
 &  & 4  & +  & -2.079  & -1.998 \tabularnewline
 & 4  & 3  & -  & -0.532  & -0.435 \tabularnewline
 &  & 4  & -  & -0.096  & -0.013 \tabularnewline
$\nu_{s}$=1, $j$=0  &  &  &  &  & \tabularnewline
 & 0  & 0  & +  & -0.095  & -0.072 \tabularnewline
\hline
\end{tabular}} 
\end{table}

The bound levels calculated in the RMA and RBAA with the present
PES are collected in Table \ref{tab:bound}. The approximate rotational
quantum number of HCN and orbital quantum number, $j$ and $l$ respectively,
are also reported in this table. The energies calculated using the
RBAA are systematically above those obtained using the RMA. This is not surprising
as the most attractive bending angle is associated with the linear
configuration of HCN. The maximum value of the total angular momentum
$J$ leading to bound states is 5 in both cases. The potential well
supports 19 bound levels and the dissociation energy is 8.986~\cm.
Harada \textit{et al.} obtained a larger dissociation energy of 9.420~\cm\
using the S02 PES, which was optimized in order to reproduce the
experimental transitions frequencies. All the bound state energies
calculated by Harada \textit{et al.} are lower than those of table II by
about a half \cm\ and they obtain one more bound state. The depth
of the S02 potential well is 30.2 \cm\ while it is 30.35 \cm\ in
the present PES. This indicates that the discrepancy in the bound
state energies does not come from the well depth, but rather from
the shape of the PES. The long-range part of the present PES may be
less attractive or its short-range part slightly more repulsive than
those of the S02 PES. 

\begin{table}[ht!!]
\caption{Comparison of observed and calculated transition frequencies in MHz. }
\centering{}\setlength{\tabcolsep}{13pt} \label{tab:exp-theo} \scalebox{0.93}{
\begin{tabular}{cccccc}
\hline 
 &  & \multicolumn{2}{c}{RMA} & \multicolumn{2}{c}{RBAA}\tabularnewline
\hline 
Transition  & Observed  & Calculated  & $\%$ error  & Calculated  & $\%$ error\tabularnewline
\hline 
j=1 $\leftarrow$ 0  &  &  &  &  & \tabularnewline
P(1)  & 97034$^{a}$  & 97696  & -0.7  & 97198  & 0,2 \tabularnewline
P(2)  & 96756$^{a}$  & 98411  & -1.7  & 97823  & 1,1 \tabularnewline
P(3)  & 98149$^{a}$  & 100188  & -2.1  & 99477  & 1,4 \tabularnewline
P(4)  & 101559$^{a}$  & 103801  & -2.2  & 102900  & 1,3 \tabularnewline
Q(1)  & 98132$^{b}$  & 101236  & -3.2  & 100534  & 2,4 \tabularnewline
Q(2)  & 101191$^{b}$  & 104373  & -3.1  & 103545  & 2,3 \tabularnewline
Q(3)  & 106244$^{b}$  & 109482  & -3.0  & 108453  & 2,1 \tabularnewline
Q(4)  & 113737$^{b}$  & 116863  & -2.7  & 115565  & 1,6 \tabularnewline
R(0)  & 98696$^{b}$  & 101006  & -2.3  & 100328  & 1,7 \tabularnewline
R(1)  & 101432$^{c}$  & 103782  & -2.3  & 102962  & 1,5 \tabularnewline
R(2)  & 105795$^{c}$  & 108350  & -2.4  & 107295  & 1,4 \tabularnewline
R(3)  & 112782$^{a}$  & 115460  & -2.4  & 114074  & 1,1 \tabularnewline
R(4)  & 122944$^{a}$  & $\cdots$  & $\cdots$  & $\cdots$  & $\cdots$ \tabularnewline
 &  &  &  &  & \tabularnewline
j=0 $\leftarrow$ 0  &  &  &  &  & \tabularnewline
R(0)  & 15894$^{c}$  & 15674  & 1.4  & 15669  & 1,4 \tabularnewline
R(1)  & 31325$^{c}$  & 30895  & 1.4  & 30892  & 1,4 \tabularnewline
 &  &  &  &  & \tabularnewline
j=1 $\leftarrow$ 1  &  &  &  &  & \tabularnewline
R(2)  & 4604$^{c}$  & 3976  & 13,6  & 3750  & 18,6 \tabularnewline
\hline
\hline 
\multicolumn{3}{l}{$^{a}$Ref. \cite{harada:02}} &  &  & \tabularnewline
\multicolumn{3}{l}{$^{b}$Average of hyperfine components from ref\cite{harada:02}.} &  &  & \tabularnewline
\multicolumn{3}{l}{$^{c}$Ref. \cite{drucker:95}.} &  &  & \tabularnewline
\end{tabular}} 
\end{table}

The calculated transitions frequencies using the RMA
and RBAA approaches are compared in Table \ref{tab:exp-theo} with
the spectroscopic data available\cite{drucker:95,harada:02}. Harada
\textit{et al.} reported most of the $Q$- and $R$-branch lines including
the splitting into several hyperfine components due to the spin angular
momentum of the nitrogen nucleus (I=1). As our calculation do not
include the hyperfine structure and because the spin splitting is
very small in comparison with the spacings of the rotational lines,
we compare our results with those of Harada \textit{et al.} averaged over the hyperfine
components. The agreement between our results and experiment is quite
good, with a difference of less than 3.2$\%$ in the RMA in all cases,
with the exception of the transition at 4604 MHz for which the error
is about 13$\%$. This is however better than the ($\sim30\%$) error
obtained by Toczylowski \textit{et al.} for this line while Harada
\textit{et al.} did not mention it. The agreement between our results
and experiment is even better when using the RBAA approach as the
maximum error is now less than 2.4\% again with the exception of
the transition at 4604 MHz for which the error is about 18.6\%. This
is the only transition which for the error is increased when using
the RBAA. 

The transition ($j$=1$\leftarrow$0)R(4) reported by Harada \textit{et
al.} is missing in our comparison as it involves the upper state ($j,l,J$)
= (1,4,5) which was not found to be bound using our PES. For each
couple ($j$=1,$l$=$n$) with $n\geq1$, there are three levels ($J$=$n$-1,$n$,$n$+1)
which are very close in energy. In the present calculation, the states
(1,4,3) and (1,4,4) have the energies -0.532 and -0.096~\cm\ respectively.
Consequently, it is not possible for the third state (1,4,5), which
is expected to be lying about $\sim$0.4~\cm\ above the state (1,4,4),
to be bound. With a potential well deeper by about a half \cm\ or
with a slightly more attractive long range interaction or less repulsive
short-range interaction, the missing state (1,4,5) could appear in
the calculations. 

\subsection{Inelastic cross sections}

\begin{figure}
\centering{\includegraphics[width=0.6\textwidth]{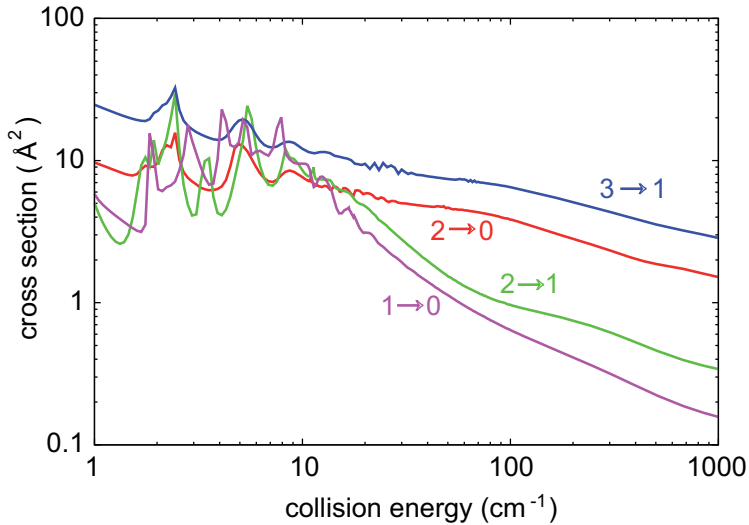}} 
\caption{\label{fig:crossRMA} Rotational transition cross sections of $l$-HCN in
collisions with He. The cross sections are given in \AA$^2$ such that the comparison with the Fig. 3 of Ref.\cite{sarrasin:10} is done readily.  }
\end{figure}

The inelastic cross sections were first calculated in the RMA in order
to compare with the previous work\cite{sarrasin:10}. Fig. \ref{fig:crossRMA}
shows the de-excitation cross sections for the first rotational levels.
The shape, the positions and the amplitudes of the resonances supported
by the van der Waals well which appear on this figure are accurate
fingerprints of the PES used in the calculations. We do not intend
here to analyse the characteristics of these resonances which are
typical of van der Walls systems and have been discussed in detail
for similar systems by several authors\cite{,balakrishnan:00}. We
simply compare our results with those of Sarrasin \textit{et al.}\cite{sarrasin:10}, 
obtained using the S01 PES. A very close similarity
is observed between the latter cross sections and the ones presented
in Fig. \ref{fig:crossRMA}, indicating that the S01 PES and the present
PES, restricted to the rigid linear HCN configuration, are very similar.

\begin{figure}
\centering{\includegraphics[width=0.6\textwidth]{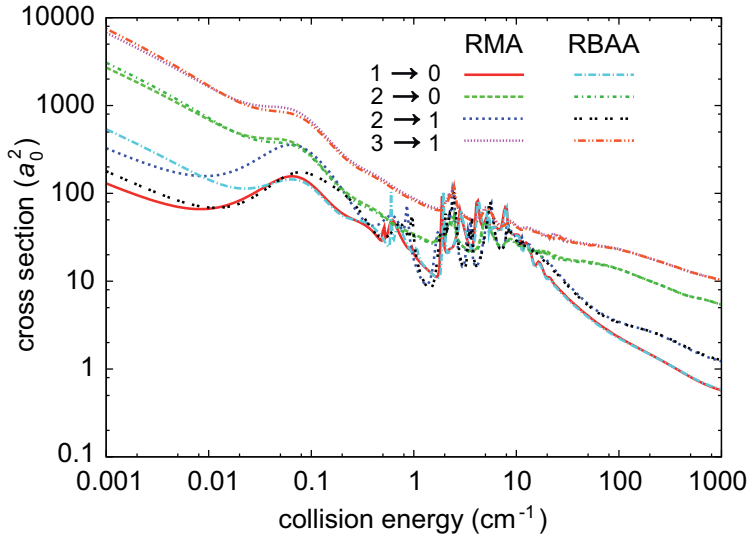}} 
\caption{\label{fig:crossRB} Comparison of the rotational transition cross sections
of HCN in collisions with He calculated
using the RMA and the RBAA approaches. }
\end{figure}

Then we investigated the bending dependance of the cross section with the present PES, by
computing the same rotational transitions using the RBAA approach. This
approach does not include exactly the coupling between vibration and
rotation which will be the object of a future work but allows checking
significant variations of the PES as a function of the bending angle.
These results are compared to those obtained using the RMA approach.
As it can be seen in Fig. \ref{fig:crossRB}, the two approaches
give very similar results. The elastic cross sections which are not
represented are almost unchanged while the inelastic cross sections
are only slightly modified at very low collision energy and around
the resonances. These very small changes show that the linear approach
is quite satisfactory to calculate rotational excitation cross sections
for a linear triatomic molecule like HCN which bending vibration frequency\cite{Maki:00}
is relatively small (711.98 $cm^{-1}$) but still large
compared to the rotational constant\cite{herzberg} (1.47822 $cm^{-1}$).

\section{Conclusion}

We presented the first theoretical study of the collision of HCN
with He including the bending vibration of HCN. We calculated a four
dimensional analytical representation of the PES based on supermolecular
\textit{ab initio} calculations using a quadruple zeta basis set with
mid-bond functions and BSSE correction. The van der Waals well was
found to be 30.35~cm$^{-1}$ deep and associated with the linear configuration
(He--HCN) while a secondary minimum with a depth of 22.08~cm$^{-1}$, associated
with a bent configuration, was also identified. Bound states calculation
were performed using this PES. The results are in good
agreement with the available experimental data.
We checked that the restriction of the dynamics to the
rigid linear configuration of HCN gives similar close coupling inelastic
cross section than the previous theoretical works. We also presented a
simple method (RBAA) of calculation of the rotational close coupling cross
section which uses the average of the interaction potential over the bending wave functions
of HCN. We found that taking into account the bending
motion through the RBAA method does not change significantly the
rotational excitation cross sections, while the agreement of the calculated bound state transition frequencies with the experiment is marginally improved.
This first study shows in any case that the RMA approach
is quite satisfactory for the computation of rotational excitation cross sections
for a linear triatomic molecule like HCN. The same accuracy could be also expected for other rigid or semi-rigid triatomic (and larger) molecules discovered in the interstellar medium. This finding is particularly useful if we consider the calculations of rotational transitions of polyatomic molecules in collision with H$_2$ which are very computationally demanding. 
Nevertheless, this preliminary conclusion needs to be confirmed by a comparison of the RMA approach with accurate calculations using an Hamiltonian which includes the exact vibrotational coupling. Efforts in that direction are in progress.

\section*{Acknowledgments}
Computer time for this study was provided by the {\it M\'{e}socentre de Calcul Intensif Aquitain} (MCIA) computing facilities of the {\it Universit\'{e} de Bordeaux} and {\it Universit\'{e} de Pau et des Pays de l'Adour}.

\bibliography{biblio_v002}

\end{document}